\documentclass[12pt]{article}

\usepackage{graphicx}
\usepackage{epsfig}
\begin{document}

\title{Propagation of Cascades in Complex Networks: From Supply Chains to Food Webs}

\author{Reginald D. Smith \\ PO Box 10051, Rochester, NY 14610 \\ rsmith@bouchet-franklin.org}

\maketitle 

\begin{abstract}
A general theory of top-down cascades in complex networks is described which explains two similar types of perturbation amplifications in the complex networks of business supply chains (the `bullwhip effect') and ecological food webs ('trophic cascades'). The dependence of the strength of the effects on the interaction strength and covariance in the dynamics as well as the graph structure allows both explanation and prediction of widely recognized effects in each type of system. \\ Keywords: Complex network; bullwhip effect; trophic cascade; supply chain; ecology\\14 pages, 3 figures
\end{abstract}
\newpage
\section{Introduction}
Complex networks permeate our lives, affecting both the natural and man-made world. Different types of networks have been studied for almost a century, however, only in the last couple of decades has complex networks emerged as a separate discipline cutting across interdisciplinary boundaries and affecting our views of systems ranging from the Internet to food webs to social networks such as instant messaging or Facebook. Much of the prominent work in this field has featured research on the evolution and properties of networks using topological measurements.
The next frontiers and pressing questions on complex networks lie with analyzing dynamics on complex networks and their relationships with topology. This is a very difficult and sometimes contentious question. For one, the growth and development of the topology and dynamics, while undoubtedly linked, usually occur on vastly separated timescales with fluctuations of the internal dynamics changing rapidly over short timescales, though perhaps with long-term trends, and with the topology often growing and developing over timescales much longer than those influenced by the key dynamic drivers.
This paper will make a contribution to this discussion by explaining how complex networks which allow flows of conserved quantities can propagate large cascades and fluctuations across the network from relatively modest perturbations at other nodes. This similar effect has been noted in several disciplines, but a unifying description of the underlying cause of this behavior amongst all types of networks has been absent. Here we will discuss two relatively well-studied examples, the bullwhip effect in supply chains and trophic cascades in food webs.

\begin{figure}[h]
\includegraphics[width=5.0in,height=3.0in]{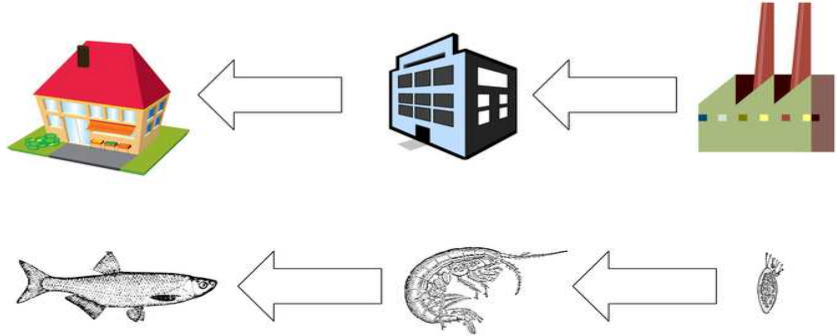}
\caption{An illustration of the two cascades discussed in this paper. Top is a supply chain whose demand is amplified from the store the factory in the `bullwhip effect'. Bottom is a simple food chain with planktivorous fish such as anchovy or mackerel consuming zooplankton which feed on phytoplankton primary producers. Changes in the population of the fish can cause a trophic cascade. Perturbations are amplified from left to right.}
\label{overview}
\end{figure}

Cascades are defined in this article as perturbations between and across different nodes in the complex network and are not identical to cascading network failures due to load and topology discussed in other literature \cite{cascade}.The first cascade, and the one most well-known amongst biological scientists, is the trophic cascade, a concept that has been discussed and debated for over forty years \cite{trophic1,trophic2,trophic3,trophic4,trophic5,trophic6,trophic7,trophic8,trophic9}. Trophic cascades are the term given for top-down effects in food webs where changes in the population of top-level carnivores will cascade through the food web, eventually causing large changes in the population of primary producers at the bottom of the web. A key example is an aquatic ecosystem such as a lake or the Black Sea \cite{trophic10,trophic11,trophic12}.   For example, a top-level predator, such as a planktivorous fish, increases their population. This causes a decline in the zooplankton population due to predation. This depression in zooplankton numbers reduces predation on phytoplankton whose population then increases. Typically, the magnitude of fluctuations at the lower primary producer level are much larger than the initial changes at the top predator level.

Trophic cascades, though recognized to exist, have been a controversial concept. They have usually been found in aquatic environments rather than terrestrial ones and seem to be most prominent where there are relatively simple linear food chains, and there is a high coupling between predator and prey with few complexities like omnivory or alternate food sources. In general, the stronger direct relationship between a single predator and prey species, the more likely a trophic cascade is to occur.

\begin{figure}[h]
\centering
\includegraphics[width=5.0in,height=3.0in]{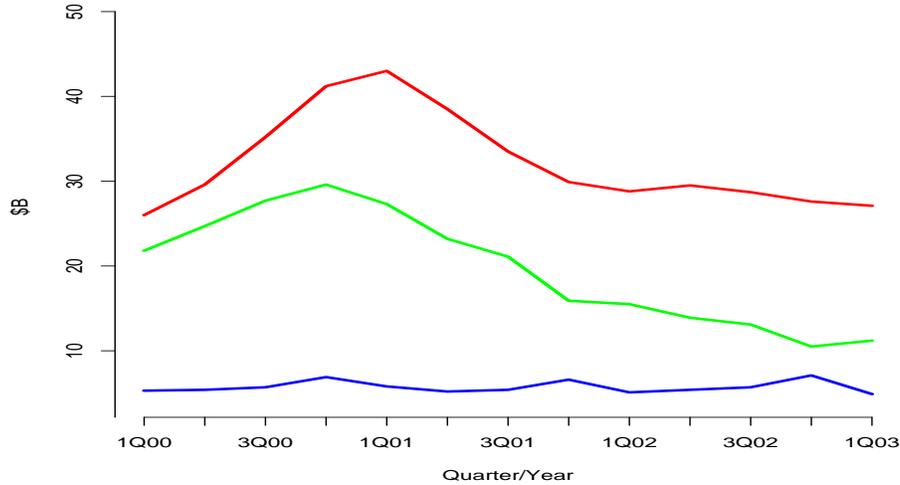}
\caption{A graph illustrating the bullwhip effect based on industry inventories at three levels of the computer communications equipment supply chain from first quarter 2000 to first quarter 2003. Blue are systems distributors, green are equipment OEMs, and red are semiconductor and electronic components manufacturers and distributors \cite{inventory}. Values are in billions of dollars. Covariances over the three years between Systems distributors and OEMs are 0.06 while between OEMs and components/semiconductors are 30.}
\label{bullwhippic}
\end{figure}

The recognition of cascades in supply chain networks was first explicitly recognized over forty years ago by researcher Jay Forrester in his treatise Industrial Dynamics \cite{forrester}, thus it is sometimes dubbed the `Forrester Effect'. Using simulations of a simple supply chain from customer to manufacturing, he showed that small changes in customer demand were successively amplified down the chain to the factory. Similar results were shown by Burbidge \cite{bullwhip1}. This effect has only recently been empirically confirmed due to the recent advances of computers and electronic visibility throughout multiple levels of the supply chain as well as new strategies of supplier collaboration and partnership. The more common term, the bullwhip effect, was first dubbed by analysts at the US consumer goods conglomerate Proctor \& Gamble (P\&G) analyzing orders for their popular diaper line, Pampers \cite{bullwhip2}. Logisticians were perplexed how fluctuations in sales at Pampers in retail stores were relatively mild, however, fluctuations at distributors were larger, causing larger scale fluctuations on orders to P\&G. By the time they analyzed orders of P\&G to its suppliers, the variances were orders of magnitude larger than those caused by end consumers. This property has since been recognized in many supply chains by Sterman and others  \cite{sterman,bullwhip3,bullwhip4,bullwhip5}. It has also been studied by physicists as a possible example of  an unstable perturbation in a system modeled using transport phenomena equations \cite{bullwhip6,bullwhip7,bullwhip8}. 
 This effect is counterintuitive and damaging to the overall level of inventory in supply chains. Low-level suppliers of parts or raw materials can often be hit by huge swings in inventory that can damage and bankrupt them. During the beginning of the global crises in 2009 when the semiconductor industry was hit hard, Morris Chang, Chairman of Taiwan Semiconductor Manufacturing Company (TSMC) lamented to the Wall Street Journal the rapid collapse of semiconductor demand had dumped inventory on those small companies in the back of the supply chain-those least able to afford it. ``Usually the guy at the rearmost end suffers the most,'' he remarked describing the damages wrought by the bullwhip effect \cite{bullwhip9}.
Like in trophic cascades, certain conditions allow the bullwhip effect to propagate more easily than others. For example, lack of supply chain visibility between suppliers and customers. When suppliers cannot reliably forecast future orders to set up purchases and balance inventory, they can overreact to small changes to assure service levels. This is exacerbated by long lead-times between ordering and delivery which demand more inventory for unanticipated demand. Also, large varieties of products, fluctuations in market pricing, and poor supplier/customer coordination can exacerbate the cascade.
Much of this is similar to trophic cascades and similar conditions and their effects are outlined in Figure \ref{overview}. However, though key aspects such as linear chains of consumption can help strengthen the cascade, there are key differences which would seem to argue the two types of cascades are of a different nature. First, trophic cascades have alternating effects between each level of the chain. An increase in predators causes a decrease in herbivores and an increase in primary producers. By contrast, in supply chains, the effect is monotonically increasingly volatile purchases and inventory levels at each lower step of the chain. In addition, ecologists have usually found strong coupling between predator and prey species strengthens the trophic cascade while strong cooperation between suppliers and customers often helps smooth out the bullwhip effect. In the next section and discussion, it will be shown that these opposite effects are not contradictory but are part of the same phenomenon which manifests itself differently in the two types of network dynamics.

\begin{figure}[h]
\centering
\includegraphics[width=5.0in,height=3.0in]{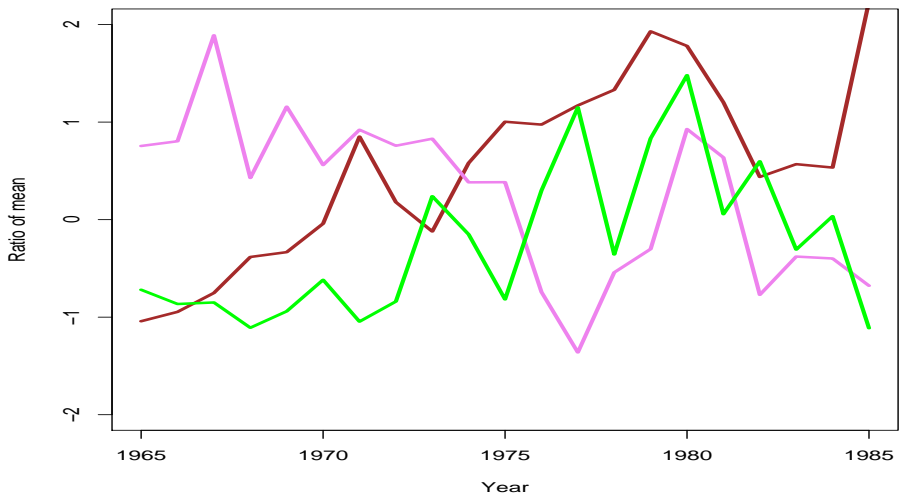}
\caption{A graph illustrating of a trophic cascade, here represented by changes in stocks of planktivorous fish (brown), zooplankton (violet), and phytoplankton (green) in the Black Sea \cite{trophicgraph}. Levels are based on ratio to the mean for each species over a period from 1950 to 2001 (only 1965-1985 shown). Covariances over the 20 year time period between the planktivorous fish and zooplankton are  -0.4 and for zooplankton and phytoplankton -0.3.}
\label{trophicpic}
\end{figure}

\section{Theoretical analysis of network cascades}
Here we consider a directed complex network, often times a hierarchical network such as a tree, where some conserved quantity, be it energy, money, or parts, flows from one set of nodes, often lower levels, to another set of nodes, often higher levels, in discrete quantities. In particular, each higher (consuming) level of the complex network has a higher density per unit of the flow quantity. Therefore, higher level predators have a higher energy/biomass density per unit (organism) than lower level herbivores or primary producers. Higher level consumers in the supply chain value their goods at a higher value per unit than lower levels due to combining inputs and adding value at each step but more relevantly, they use multiple units of input to produce one unit of output. Also, high level suppliers tend to purchase individual products in larger batch sizes, with the exception of individual consumers. Here we will consider the interaction strengths between higher and lower levels by the variable, $a_{ij}$. Where the interaction strength represents for every unit of  $i$, the higher level/consuming entity, $a_{ij}$ units of $j$, the lower level, are consumed. This roughly correlates with predator-prey interaction strengths in biology or supply quantities based off bills of materials or batch order sizes in supply chains. 

The growth rate of each node is given by a simple birth/death process where 

\begin{equation}
G = \frac{dP_n}{dt}= B-D
\end{equation}

where $G$ is the growth rate, $P_n$ is the unit amount at level $n$, $B$ is the `birth rate' or rate of unit growth and $D$ is the `death rate' or decrease in unit growth. This represents the inventory growth for each level in a supply chain network or the per capita ($dP/P$) growth in the populations of food webs. The `death rate' in natural populations is consumption by predators, while is supply chain networks it is the purchase of inventories by customers. The birth rate is the Malthusian growth rate for a natural population or the purchase of supplies to replenish inventories by a company in a supply chain.
In order to investigate the cascades effectively, the key important variable will not be $P_n$ or even $G$ but rather the variance of $G$ at a node. This can be generally given at node $j$ by

\begin{equation}
\sigma_j^2 = \sum^n_{i=1}a_{ij}^2\sigma^2_{Bi} - a_{ij}Cov(B_i,B_j)
\end{equation}

where $a_{ii}=1$ and $Cov(B_i,B_i) = \sigma_i^2$. Therefore, it is clearly seen then that the variance at a node is effectively due only to interactions with other nodes since the effects of its own variance and self-covariance cancel out. In addition, the three factors that integrate to determine the strength of the cascade are the variance, covariance, and interaction strength.
Now we will look at how this plays out in a network where we will look at how these three factors integrate with the graph structure of a network to determine the overall variance at a given node.
Define the adjacency matrix of a digraph as $A$ where $A_{ij}$ is 1 where there is a directed edge between two nodes and 0 otherwise. Also, define the principle eigenvalue, $\lambda_1$ of $A$ as the largest eigenvalue of the eigenspectrum of $A$. Our first analyses will occur on the restricted example of a regular digraph, in other words, a graph where every node has the same in-degree $k_{in}$. According to spectral graph theory \cite{spectral}, if the graph is strongly connected and any node is accessible from any other node, $\lambda_1$ and $k_{in}$ are related as

\begin{equation}
\lambda_1 = k_{in}
\end{equation}

Now, regarding the interaction strengths and covariances, let us make the restrictive assumption that there is an average interaction strength between all connected nodes, $\bar{a}$ and an average variance $\bar{\sigma}^2$ and covariance $\bar{\sigma}_{ij}$. The variance at every node in the network, using the multiplicity factor $k_{in}$, can be defined as

\begin{equation}
\sigma^2_i = k_{in}\bar{a}(\bar{a}\bar{\sigma}^2 - \bar{\sigma}_{ij})
\end{equation}

or

\begin{equation}
\sigma^2_i = \bar{a}\lambda_1(\bar{a}\bar{\sigma}^2 - \bar{\sigma}_{ij})
\end{equation}

Now, assume that the adjacency matrix can also be regarded as an interaction matrix. For a typical adjacency matrix, this would define $\bar{a} = 1$. However, from spectral graph theory it is also realized that when you multiply $A$ by a constant such as $cA$, then all eigenvalues are also multiplied by the same and become $c\lambda$. So for $\bar{a}A$ we can simplify the previous equation for any interaction strength as

\begin{equation}
\sigma^2_i = \lambda_1(\bar{a}\bar{\sigma}^2 - \bar{\sigma}_{ij})
\label{eigeneq}
\end{equation}

From these equations we can see a simple fact that for a regular graph with a mean interaction strength, the variance can be largely predicted based off of the principal  eigenvalue which consolidates the structure and interaction dynamics into one metric. It also helps to illuminate one of the main interests of this paper: namely, why the behavior of trophic cascades and the bullwhip effect differ. Namely, the key factor is the subtraction of the mean covariance. For systems with positive covariance, such as supply chain networks, a higher covariance which indicates greater supply chain coordination will reduce the overall variance and damp the bullwhip effect. For systems with negative covariance, like most interactions between trophic levels, greater coupling between predator and prey actually increases the overall variance of the system. Therefore, we can see the two phenomena are based on a similar underlying mechanism.

We can expand our model, however, since it is a poor approximation to most real systems. The interaction strength in no real network is constant across all connected nodes, much less the variance and covariance. For a first order more realistic approximation, let us look at a more general graph. This time, we still keep a mean variance and covariance but we assume that the interaction strength between one node and all other nodes is the same inversely proportion to the in-degree so that 

\begin{equation}
a_{ij} = \frac{a_{max}}{k_i}
\end{equation}

Thus the variance would become

\begin{equation}
\sigma^2_i =\frac{a_{max}^2}{k_i^2}\bar{\sigma}^2 - \frac{a_{max}}{k_i}\bar{\sigma}_{ij}
\end{equation}

Thus, the interaction strength plays a dominant role while the in-degree acts to damp the contribution of variance from other nodes. If $k_i \gg a_{max}$ for hub nodes we can see a situation where the variance of the node is completely damped out.

In these two examples, we see the general effect of the graph structure and interaction strength on the variance of a node, but one of the key questions in cascades is how the variance of one node ultimately effects another since a key feature of dynamics like the bullwhip effect is the magnification of small variance across levels of the supply chain.
For two connected nodes $i$ and $j$, the effect of $i$ on $j$ is simple

\begin{equation}
\sigma^2_j =a_{ij}^2\sigma_i^2 - a_{ij}\sigma_{ij}
\end{equation}

Now what about the effect of $i$ on $k$ who are not directly connected by separated through $j$?
\begin{eqnarray}
\sigma^2_k =a_{jk}^2\sigma_j^2 - a_{jk}\sigma_{jk} \nonumber \\
\sigma^2_k =a_{jk}^2 \big(a_{ij}^2\sigma_i^2 - a_{ij}\sigma_{ij}\big)- a_{jk}\sigma_{jk} \nonumber \\
\sigma^2_k = a_{ij}^2 a_{jk}^2\sigma_i^2 - a_{jk}^2 a_{ij}\sigma_{ij} - a_{jk}\sigma_{jk} \nonumber \\
\end{eqnarray}
Only the last term does not incorporate interactions from node $i$ so in general the effect of any node on another (node $n$) through a path $1\dots n$ is given by

\begin{equation}
\sigma^2_n = \sigma_1^2 \prod_i^{n-1} a_{i,i+1}^2 - a_{12}\sigma_{12}\prod_{i=2}^{n-1} a_{i,i+1}^2
\label{cascadestrength}
\end{equation}

Therefore, the total effect is the product of square of the interactions along the path between the two nodes times the variance of the starting node with roughly the same effect on the covariance. This helps to explain the massive amplifications in variance across relatively few levels of a network. If you assume that the interaction scales by some constant amount, say $\beta$ at each step and each node is separated an average length $\bar{l}$ the average effect can be estimated as
\begin{equation}
\sigma^2_n = \beta^{2\bar{l}}(\sigma_1^2 - \sigma_{12})
\end{equation}
If we always assume $\beta > 1$ for increasing density at different levels this shows we clearly have an exponential effect. 
In one last example, we will look at the example of a heterogeneous network be it exponential, scale-free, or what not with an assortative distribution of in-degrees $P(k_1,k_2)$. Making the assumption that the interaction strength between two nodes depends on their relative degrees, we define $a(i,j)$ as
\begin{equation}
a(k_i,k_j) = a\frac{k_i}{k_j}
\label{SFstrength}
\end{equation}
So if $i$ has a higher degree than $j$ the coupling is large or vice versa if the edge is directed from a small edge node to a larger edged one. Using equations \ref{cascadestrength} and \ref{SFstrength} we can clearly see that overall the effect of a cascade chain from 1 to $n$ where the average path length is $\bar{l}$ is approximately

\begin{equation}
\sigma^2_n = a^{2\bar{l}}\bigg( \frac{k_1}{k_n} \bigg)^2(\sigma_1^2 - \sigma_{12})
\end{equation}
In other words, though the interaction strength constant is raised to the exponential of twice the path length, only the degrees of the first and final node matter as a ratio for modifying this. Thus, looking at the effect of a low-degree node from anywhere on the network we can see that the interaction could be highly muted by the large degree of the destination node. However, in the opposite, a cascade initiated from a high degree node could have serious reverberations throughout the network with a hugely magnified effect at smaller nodes. Obviously, this depends on the validity of the interaction strength being proportional to the degrees of origin and destination nodes.
As a final note, we should consider that one other theorem from spectral graph theory states that for a mean degree $\langle k_{in}\rangle$ and principle eigenvalue $\lambda_1$ for any graph

\begin{equation}
\langle k_{in}\rangle \leq \lambda_1
\end{equation}

with equality only for regular graphs. Therefore in a graph with an identical mean interaction strength (where it can be represented as the adjacency matrix times a constant) equation \ref{eigeneq} represents the maximum possible node variance and this is much reduced for graphs that have the same average degree of an equivalent regular graph but are more heterogeneous. Therefore, in this restricted situation, more complex topologies actually help damp the variance overall at nodes and reduce the strength of the cascades. For linear chains and trees without loops, despite having an eigenspectrum where all eigenvalues are 0, with an average mean interaction strength between nodes, they demonstrate the same coupling and behavior as a regular graph with $\langle k \rangle = 1$.

\section{Discussion}
The examples above show that in graphs, the structure and the interaction strength can have a subtle and complex interplay in determining how the variance of one node effects the other in its neighborhood or even in another part of the graph. In general we can see several constants though. For identical interaction strengths across all nodes, the cascades are increased by more dense graphs (more in-degrees) and topological similarity to regular graphs (less heterogeneity). When the interaction strength depends on degree, the interaction strength alone takes a greater role in determining cascade strength, especially in high degree nodes. Finally, for interaction strengths depending on the relative degrees of two nodes, propagation is largely determined by the relative degrees of the beginning and ending node.

The effect of the interaction strength produces a multiplicative cascade as it travels through the network in how it modifies both the variance at each step. The key to the difference in cascades in supply chains and food webs is the nature of the covariance between the nodes. In supply chain networks, growth of connected firms is essentially a positive feedback loop and thus positively correlated. Sales in a consumer tend to drive sales at a supplier. Therefore, the covariance is generally positive. Given the minus sign in front of the covariance term, increased covariance will tend to dampen the fluctuations in all nodes of the supply chain. This is already well-known in practice. By increasing supply chain visibility and cooperation, usually through forecasts and optimized lot sizes/deliveries combined with production and Lean initiatives, one can theoretically increase the covariance between the production of the two firms, which would reduce the overall variance of inventory levels.

The opposite effect is seen in food webs because the relationship between the growth of predator and prey populations is typically an inverse one and thus has negative feedback and thus negative covariance. The importance of negative covariance in the growth of populations in food webs has been long recognized and is a key theory in the diversity-promotes-stability debate in ecology \cite{diversity}. Growth in a predator population reduces the prey population and its growth rate. Since predator-prey relationships have a negative covariance, tighter coupling, which implies a larger absolute covariance but a more negative value, actually increases the instability and variance at the lower levels. It has long been known that in ecosystems, the ultimate effect of trophic cascades on primary producers is based on whether the number of trophic levels is even or odd \cite{trophic14}. This is why decoupling of predator and prey species through activities like omnivory tends to dampen rather than increase the effects of trophic cascades unlike in supply chains where lack of collaboration can lead to volatile inventory fluctuations and market oversupply. The coupling of interactions amongst complex food webs can be very useful in explaining observed phenomena or possibly finding out to prevent large-scale extinction cascades \cite{extinct}.

Though the theoretical insights in this paper may serve as general guides, reality is always more complicated. First, interaction strengths, especially in food webs, tend to be heavily skewed where a few key interactions dominate while many small interactions have much less strength \cite{trophic15,trophic16,trophic17}. Second, research in complex networks tends to show that degree distributions are also skewed and long-tailed, therefore though an average description can give us theoretical insight into the impact of network parameters and interaction strengths, we cannot ignore the heterogeneity in the network by which some units or subsets of the network are subject to greater pressures (economic or ecological) and thus may show markedly different behavior than the entire network. Given that the skewed nature of degree distributions dictates most nodes have only a few edges, a large average in-degree may only represent a relatively smaller number of highly interacting units in the network.  

Finally, there needs to be a greater description of changing and coupling of interaction strengths and covariances due to feedback effects. Under a static model, a node could feasible provide huge feedback on its own variance through cycles in the network where they exist. These cycles would have to either have a weak interaction overall or negative feedback moderation to prevent runaway variance effects.

The birth and death variables used here can be expanded given a specific system to include exponential growth, nonlinear interactions, and time lags that characterize real systems. In fact, these systems of equation are easily amenable to Lotka-Volterra type interactions or simple continuous inventory ordering systems. The general amplification of distortions likely applies to a variety of known and yet unknown systems. As a final note, though the introduction of the network and interaction parameters may seem to imply a message for the diversity versus stability of an interaction network, it is not completely clear on this point. It does not consider general environmental effects and should be carefully differentiated from overall stability to perturbations or extinctions which characterize most of the broader questions of ecosystem stability primarily addressed by the debates. However, it does seem that in order to understand the full effect of diversity in network one must take into account both the general network, the dynamics, and the covariance amongst population growth rates over time to fully understand observed behavior.

Acknowledgements: The author would like to acknowledge the gracious advice and help of Dr. Lawrence Thomas. I also thank Dr. Georgi Daskalov of CEFAS Lowestoft Laboratory for providing the trophic cascade data from his Black Sea paper for Figure 3.

\end{document}